\documentclass[10pt]{article}
\pdfoutput=1
\usepackage{graphicx,amssymb,amsfonts,latexsym,times,amsmath,amsthm}

\usepackage{vmargin,nccmath,enumitem}

\usepackage{graphics,float,indentfirst,authblk}

\usepackage[english]{babel}
\usepackage{color}

\title{Optimal vaccine allocation during the mumps outbreak in two SIR centers}
\author{A. Chernov\thanks{Corresponding author:
Alexey Chernov, National Research University Higher School of Economics, Moscow, Russia;
E-mail: aachernov@hse.ru} $^{1}$,
M. Kelbert$^{1,2}$
and A. Shemendyuk$^{1}$ \\
 $^{1}$ National Research University Higher School of Economics, Moscow, Russia\\
 $^{2}$ Swansea University, Singleton Park, SA2 8PP, United Kingdom}

\begin{document}
\maketitle
\begin{abstract}
The aim of this work is to investigate the optimal vaccine sharing between two SIR centers in the presence of migration fluxes of susceptibles and infected individuals during the mumps outbreak. Optimality of the vaccine allocation means the minimization of the total number of lost working days during the whole period of epidemic outbreak $[0,t_f]$, which can be described by the functional $Q=\int_0^{t_f}I(t){\rm d}t$ where $I(t)$ stands for the number of infectives at time $t$.
We explain the behavior of the optimal allocation, which depends on the model parameters and the amount of vaccine available $V$.
\end{abstract}
\providecommand{\keywords}[1]{\textbf{{Keywords:}} #1}

\quad 

\section{Introduction}

Epidemics are one of the most destructive and dangerous phenomena for humanity.
There have been numerous historical confirmations of terrible pandemics that devastated vast territories and destroyed millions of people. Such epidemics are also possible now. With the development of mathematical models, such as the Kermack-McKendrick or SIR model (see, e.g.,~\cite{l1,l2,l3}), it has become possible to predict spread of epidemics over time (number of susceptible, invectives and immunized).

In this paper we consider the epidemics of mumps and optimal vaccine allocation for it. Mumps is a viral disease caused by the mumps virus. Currently many countries are experiencing mumps outbreak. For example, one of the largest mumps outbreaks in the United States was in 2006. Occurring in January, state Iowa, the virus had quickly spread through 40 states by April resulting in 2786 reported cases throughout the country. In December of the same year the total number of cases had increased to 6584 with 85 hospitalizations and no deaths ~\cite{m1}.

The aim of this work is to investigate an optimal vaccine allocation in two SIR centers connected by the steady migration fluxes during the mumps epidemics.
The problem of optimal vaccination has been actively researched since 1970s (see pioneer works \cite{n1,n2} and
more recent \cite{n3,n4,n5,n6,n7,n8,n9,n10,n11,n12,n13,n14, l4, l5}). 
Most of these works concentrate on an isolated center and abundance of vaccine that is enough to cover a significant fraction of the population. Unlimited amount and a quadratic cost of vaccines were considered in the case of $n$ centers in~\cite{l6}.

Limited amount of vaccine and its optimal allocation were investigated for non-interacting centers in deterministic ~\cite{l7} and stochastic ~\cite{l8} cases. Interacting centers without migration in deterministic and stochastic cases were considered in~\cite{l9}. The optimal vaccine allocation in different vaccination schedules in two centers was
studied  in ~\cite{n15} for the typical fixed parameters values without analyzing the dependence of the parameters domain-wide.
Interacting centers in the form of epidemic percolation network were considered in~\cite{n16,n17}.  In \cite{l7} authors maximize the total number of people that escape infection, in papers~\cite{l8,l9} they minimize the final quantity of removed and infectives, in paper~\cite{n15} they minimize a mortality. These functionals are interesting from a healthcare point of view. Unlike previous articles, we concentrate on minimizing  the natural functional $Q = \int_0^{t_f} I(t) dt$,  where $I(t)$ is the number of infectives and $t_f$ is the final moment of epidemic:  $t \in [0,t_f]$. Moreover, we explore the dependence of $Q$ from the parameters of the model. This functional is interesting both in healthcare and medical insurance since it relates to the total number of lost working days for the entire period of epidemic. To make our model more plausible, we added migration fluxes  between centers.

The classical SIR model in case of $n$ centers and minimization of functional $Q$ is governed by the following ODEs (see e.g.~\cite{l11}):
\begin{equation}\label{u1}
\begin{aligned}
  \frac{dS_i}{dt}&=-\frac{\beta}{N_i} S_i I_i - \sum_{j \neq i} k_{ij} S_{i}+\sum_{j\neq i} k_{ji} S_j - \min(w_i V , S_i) \delta(t-t^*)  \\
  \frac{dI_i}{dt}& = \frac{\beta}{N_i} S_i I_i - \alpha I_i-\sum_{j \neq i} l_{ij} I_i(t)+\sum_{j\neq i} l_{ji} I_j  \\
  \frac{dR_i}{dt} &= \alpha I_i(t) + \min(w_i V, S_i) \delta(t-t^*)\\
  Q(w_1, w_2, \ldots, w_n)&=\sum_{i=1}^n \int_0^{t_f}  I_i(t) {\rm d}t \rightarrow \min_{(w_1, w_2, \ldots, w_n): w_1+w_2 + \ldots + w_n = 1}\\
  S_i(0)&=S_{i0}, I_i(0)=I_{i0},
  R_i(0)=R_{i0}\\
  \sum_{i=1}^n  [S_i(t) &+ I_i(t) + R_i(t)] = \sum_{i=1}^n  N_i \\
  V &\leq \sum_{i=1}^{n}{S_i(0)}.
 \end{aligned}
\end{equation}
Here  $S_i(t),I_i(t),R_i(t)$ are interpreted as numbers of susceptibles, infectives and removed in the $i$th center respectively,
$V$ is amount of vaccine available,
$w_i$ is the fraction of vaccine stock $V$ allocated in the $i$th center,
$t^*$ is the vaccination time,
$\alpha$ is the recovery rate,
$\beta$ is the infection rate,
$k_{ij}$ is the transportation rate of susceptibles from center $i$ to center $j$,
$l_{ij}$ is the transportation rate of infectives from center $i$ to center $j$,
$i, j = 1, \ldots ,n$,
$\delta(t-t^*)$ is Dirac's delta function implying a unit impulsive increase in the number of vaccinated individuals.
Note that in classic SIR model the population remains constant. Therefore, we abandon the removed group $R$, since it can be deduced from $S$ and $I$.

Rather than prophylactic or proactive vaccination, we focus on reactive vaccination administered after the initial disease outbreak.
The common wisdom suggests an early allocation of all available resources just after the beginning of epidemic outbreak, but logistics create an unavoidable delay.
The optimal division of this vaccine crucially depends on the population sizes and migration rates.
We analyze these dependencies in a simplified model, where after a vaccination at the moment $t^*$ susceptibles immediately transform into the removed group. In numerical simulations we use \textit{odeint} function from \textit{SciPy} package in Python. However, due to Dirac's delta function in Section~2, we use explicit Euler method to derive the optimal vaccination time $t^*$.

Our paper is organized as follows. In Section~2 we account for the vaccination delay and analyze the dependence $Q$ on the total amount of vaccine in an isolated SIR center.
In Section~3 an optimization of vaccine sharing between two interacting SIR centers is studied for the cases of two identical (Subsection~3.1) and asymmetric (Subsection~3.2) SIR centers. Summary and an outlook for a further work is given in Section~4.

\section{An isolated center}\label{sec:SIR}

The model of an isolated SIR center with vaccination at the moment $t^*$ takes the form:
\begin{equation} \label{u2}
\begin{aligned}
\frac{dS}{dt} &= -{\beta \over N} S I - \min(S, V) \delta(t-t^*)  \\
\frac{dI}{dt} &= {\beta \over N} S I - \alpha I\ \\
\frac{dR}{dt} &= \alpha I + \min(S, V) \delta(t-t^*) \\
V &\leq S(0),\quad S(t) + I(t) + R(t) = N\\
Q(t^*)&=\int_0^{t_f} I(t) {\rm d}t \rightarrow \min_{t^*}\\
\end{aligned}
\end{equation}

The basic reproduction number is defined as $\mathbf{R_0}={\beta \over \alpha}$ and in case of mumps $\mathbf{R_0} =4$.
Without loss of generality we can set the initial number of removed $R_0=0$ and let $\alpha = 1$ and $\beta = 4$.

We restrict our consideration to the instantaneous vaccination, but similar analysis can be done for more realistic vaccination schedules.

\begin{figure}[H]
\centering
\hspace{-0.3\textwidth}
\includegraphics[width=0.50\textwidth]{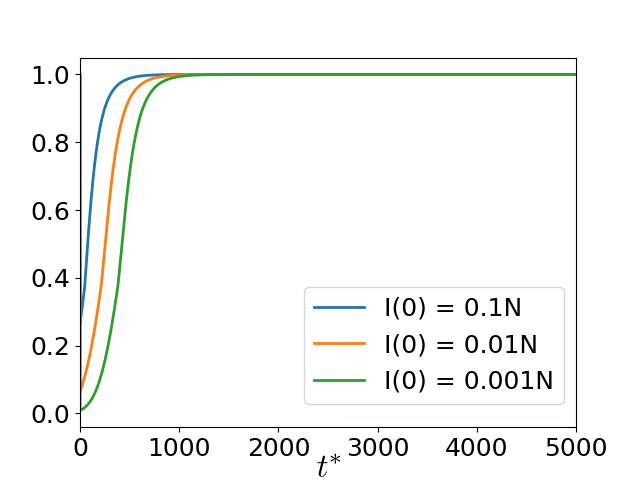}
\includegraphics[width=0.50\textwidth]{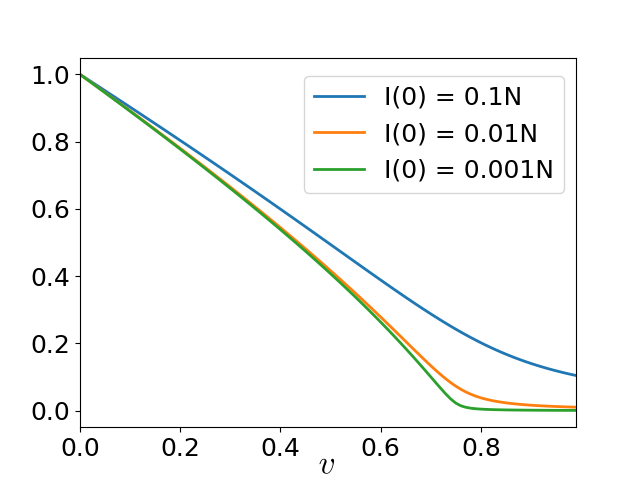}
\hspace{-0.3\textwidth}
\caption{Left:
Functional $Q(t^*)$ for different $I(0)$, normalized to the maximum value of the functional. $t^*$ is the vaccination time. Parameters: $\mathbf{R_0}=4, v=0.4$
\protect\newline
Right: Functional $Q(v)$  for different $I(0)$, normalized to the maximum value of the functional. Parameters: $\mathbf{R_0}=4, t^* = 0$.
}
\label{fig:1-2}
\end{figure}

Firstly, we analyze the dependence of $Q$ on the vaccination time for fixed share of the vaccine $v={V \over S(0)}$. Results of the numerical integration of ODEs~(\ref{u2}) plotted in Figure~\ref{fig:1-2}(left) show that the optimal time of vaccination is $t^*=0$ and the increase rate of $Q$ depends on the
initial number of infectives. Further time is measured in the number of iteration in the simulation.

The greater number of initially infectives $I(0)$ is, the more critical is the effect of latency (see Figure~\ref{fig:1-2} (left)). Thus, the optimal time of vaccination is the time of epidemic registration $t^*=0$.

Now we assume that the vaccination time is optimal $t^*=0$. It is interesting to consider the dependence of the functional $Q$ on the available vaccine stock. Dependence of the functional $Q$ on the vaccine share $v \equiv \frac{V}{S(0)}$ for different values $I(0)$ is shown in Figure~\ref{fig:1-2} (right). Note that the dependence is not linear and for any parameters there exists a threshold $v^*(I_0)$ such that $Q$ becomes sufficiently small and does not change significantly for $v \geq v^*$.
The amount $v^*$ is essentially enough to suppress the epidemic as demonstrated in Figure~\ref{fig:1-2}(right).

\section{Two interacting SIR centers}\label{sec:2SIR}

In this section we study numerically the model described by equations (\ref{u1}) with vaccination time $t^* = 0$ for the simplest network of two interacting SIR centers, and show how to allocate the available vaccine optimally depending on the model parameters.

The system of ODEs~(\ref{u1}) in case of 2 centers has the following form:

\begin{equation}\label{u3}
\begin{aligned}
  \frac{dS_1}{dt}&=-\frac{\beta_1} {N_1}  S_1 I_1 - k_{12} S_{1} + k_{21} S_2 \\
  \frac{dS_2}{dt}&=-\frac{\beta_2} {N_2} S_2 I_2 - k_{21} S_{2} + k_{12} S_1 \\
  \frac{dI_1}{dt}& = \frac{\beta_1} {N_1} S_1 I_1 - \alpha I_1- l_{12} I_1 + l_{21} I_2  \\
  \frac{dI_2}{dt}& = \frac{\beta_2} {N_2} S_2 I_2 - \alpha I_2- l_{21} I_2 + l_{12} I_1  \\
  \frac{dR_1}{dt} &= \alpha I_1\\
  \frac{dR_2}{dt} &= \alpha I_2\\
  Q(w_1, w_2)&= \int_0^{t_f} [I_1(t) + I_2(t)] \quad {\rm d}t \rightarrow \min_{(w_1, w_2): w_1+ w_2 = 1} \\
  S_j(0)&=S_{j0} - \min\left(S_{j0}, w_j  V\right)\\
  I_j(0)&=I_{j0}\\
  R_j(0)&=R_{j0} + \min\left(S_{j0}, w_j  V \right)\\
  S_{j0}&+ I_{j0} + R_{j0} = N_j, \quad j=1,2\\
  V &\leq S_1(0) + S_2(0), \quad \sum_{j=1}^2 [S_j(t) + I_j(t) + R_j(t)] = N_1 + N_2.
 \end{aligned}
\end{equation}
Clearly, the total number of vaccine is less than total number of susceptibles $V \leq S_1(0) + S_2(0)$. Further we consider vaccine share and let $v = \frac {V} {S_1(0) + S_2(0)}, 0 \leq v \leq 1.$
For example, if $v = 1$ we can vaccinate all population, if $v=0.5$ - only half of susceptibles, etc.

Note that if the first center receives a vaccine share $w_1$, then
the second center receives the rest of the fraction $w_2 = 1 - w_1$ even if the share of vaccine $w_2 V$ is greater than the share of susceptibles in the second center, which is not rational. Clearly, to minimize the functional $Q (w_1, 1 - w_1) \equiv Q (w_1)$ we need to use all of the vaccine volumes available.
However, we will neglect these situations to simplify computer calculations and also examine nonoptimal vaccine allocations.

\subsection{Identical centers}\label{sec:sym}
\begin{figure}[H]
\centering
\hspace{-0.3\textwidth}
\includegraphics[width=0.50\textwidth]{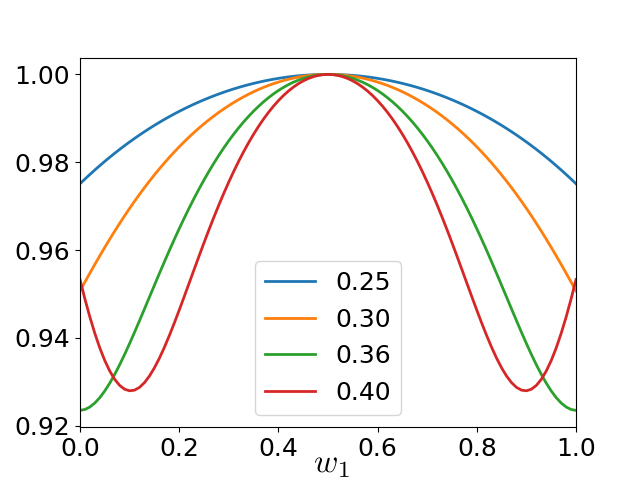}
\includegraphics[width=0.50\textwidth]{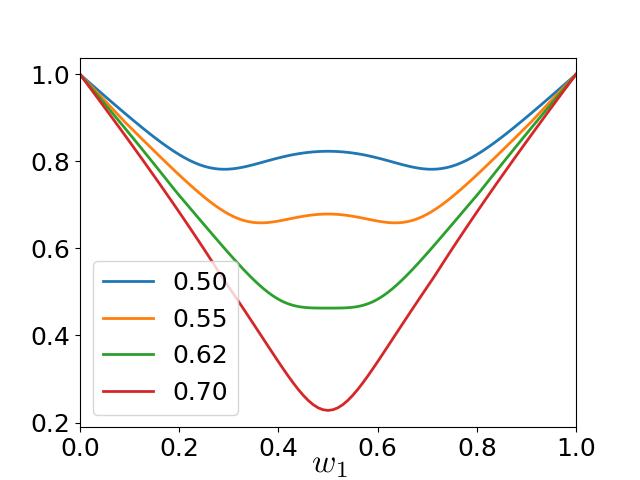}
\hspace{-0.3\textwidth}
\caption{Left: Functional $Q(w_1, w_2)$, $w_1+w_2=1$ for different $v \in [0.25,0.4]$, normalized to the maximum value of the functional. Parameters: $\mathbf{R_0}=4, k=0.01, l=0.001, I=0.01N$.\protect\newline
Right: Functional $Q(w_1, w_2)$, $w_1+w_2=1$ for different $v \in [0.5,0.7]$, normalized to the maximum value of the functional. Parameters: $\mathbf{R_0}=4, k=0.01, l=0.001, I=0.01N$.}
\label{fig:5-6}
\end{figure}

Consider two identical SIR centers connected by symmetric migration fluxes and described by ODEs  ~(\ref{u3}). We are studying the dependence of epidemic spread on allocation of the vaccine in terms of the functional $Q$ for different $v$ in $[0.25,0.70]$.
For a basic example consider the following parameters: $N_1=N_2=N=1000$, $\beta_1=\beta_2=4, \alpha=1, k_{12}=k_{21}=k=0.01, l_{12}=l_{21}=l=0.001, I_1(0)=I_2(0)=I=10$. Results of the simulations for different fractions of the vaccine $w_1$ are presented in Figure~\ref{fig:5-6}.

From Figure~\ref{fig:5-6} we conclude that there are two thresholds $v^*_1 \approx 0.36$ and $v^*_2 \approx 0.62$. If $v < v^*_1$ the minimum value of functional $Q(w_1)$ is reached at the ends of the segment. For $v > v^*_2$ he minimum value of functional $Q(w_1)$ is reached at $w_1 = 0.5$. Thus, if the vaccine share is small, we need to fully vaccinate only one of the centers (by symmetry it does not matter which center we choose); for large volumes of vaccine it is most reasonable to divide equally between the centers; otherwise, there is an optimal $w_1 \neq 0.5$ within a segment that minimizes $Q(w_1)$.

Threshold $v^*_1$ is the most interesting. With increase of the vaccine share $w_1$ we abandon the previous strategy and start to divide the vaccine according to Figure~\ref{fig:5-6} (right). An existence of the threshold $v^*_1$ may be due to the existence of the similar threshold $v^*$ in the SIR model with one center. The additional vaccine units in model with one center does not change the functional significantly. In the model with two identical centers it means that if there is a sufficient level of vaccine in one center, an additional unit of vaccine is much more valuable in another city. Eventually we divide vaccine equally between two centers.

Further, our goal is to study the dependence of the threshold on the parameters: $v^*_1=v^*_1(I, l_{ij}, k_{ij})$

\textbf{1.} \textbf{Dependence on the initial number of infectives} $I_{1,2}(0)=I$

\begin{figure}[H]
    \centering
    \hspace{-0.3\textwidth}
    \includegraphics[width=0.50\textwidth]{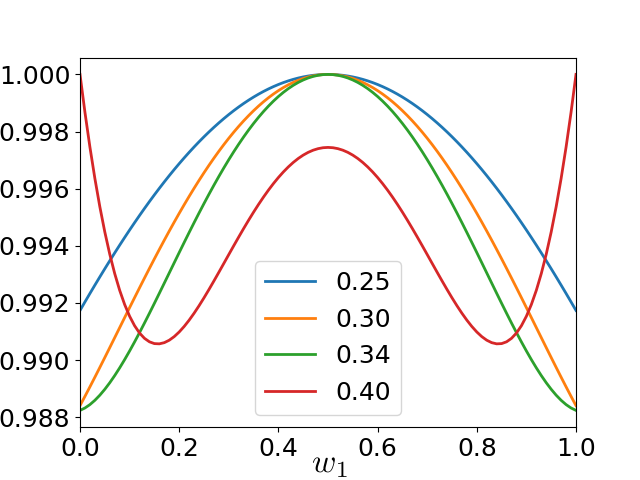}
    \includegraphics[width=0.50\textwidth]{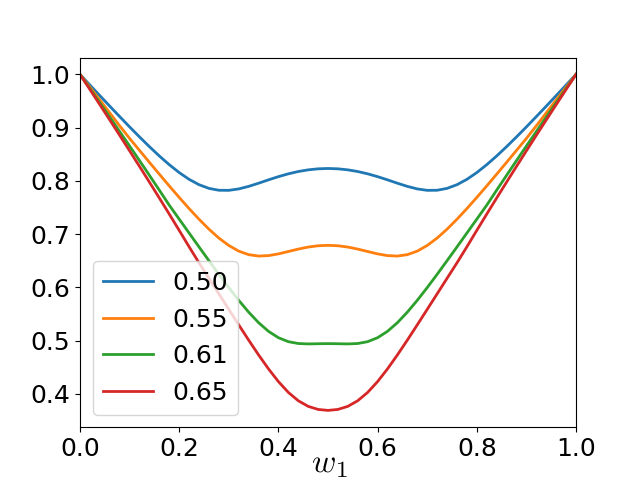}
    \hspace{-0.3\textwidth}
    \caption{Left: Functional $Q(w_1, w_2)$, $w_1+w_2=1$ for different $v \in [0.25,0.4]$, normalized to the maximum value of the functional. Parameters: $\mathbf{R_0}=4, k=0.01, l=0.001, I=0.1N$.\protect\newline
    Right: Functional $Q(w_1, w_2)$, $w_1+w_2=1$ for different $v \in [0.5,0.65]$, normalized to the maximum value of the functional. Parameters: $\mathbf{R_0}=4, k=0.01, l=0.001, I=0.1N$.
    }
    \label{fig:9-10}
    \end{figure}

Dependence of the functional $Q$ on $I$ is shown in Figure~\ref{fig:9-10}, where the initial value of infectives in both centers is 10 time higher than in the basic example (see Figure ~\ref{fig:5-6}). In this case, the threshold decreases to $v^*_1 \approx 0.344$. We need to divide the vaccine between two centers for $v \geq v^*_1$. The large number of initial infectives can be interpreted as vaccination delay. As we have seen in Figure~\ref{fig:1-2} this delay significantly reduces the effectiveness of vaccination.

\textbf{2. Dependence on the migration activity of susceptibles}

\begin{figure}[H]
    \centering
    \hspace{-0.3\textwidth}
    \includegraphics[width=0.50\textwidth]{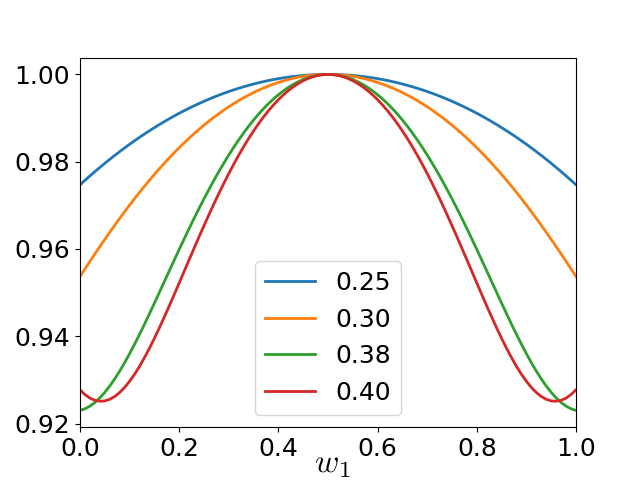}
    \includegraphics[width=0.50\textwidth]{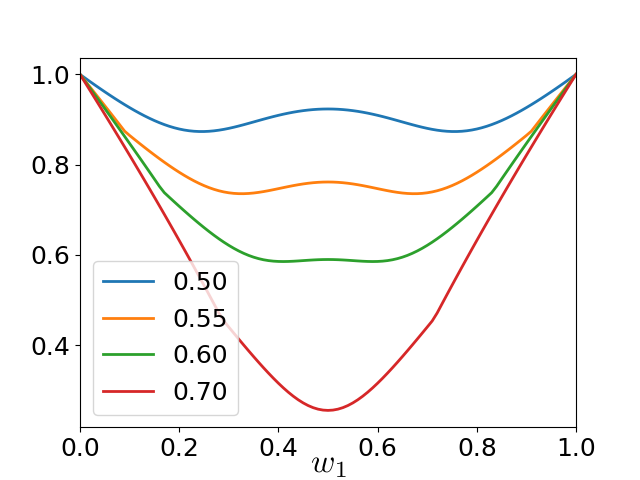}
    \hspace{-0.3\textwidth}
\caption{Left: Functional $Q(w_1,w_2)$, $w_1+w_2=1$ for different $v \in [0.25,0.4]$, normalized to the maximum value of the functional. Parameters: $\mathbf{R_0}=4, k=0.1, l=0.001, I=0.01N$.\protect\newline
    Right: Functional $Q(w_1,w_2)$, $w_1+w_2=1$ for different $v \in [0.5,0.7]$, normalized to the maximum value of the functional. Parameters: $\mathbf{R_0}=4, k=0.1, l=0.001, I=0.01N$.}
    \label{fig:11-12}
\end{figure}

Let the migration activity of susceptibles $k_{ij}$ increase 10 times in both centers.
Then the threshold increases to $v^*_1 \approx 0.384$. When $v > v^*_1$
and the migration activity of healthy people is high,
it becomes more efficient to allocate the vaccine in both centers (see Figure~\ref{fig:11-12}).

\textbf{ 3. Dependence on the migration activity of infectives}

\begin{figure}[H]
    \centering
    \hspace{-0.3\textwidth}
    \includegraphics[width=0.50\textwidth]{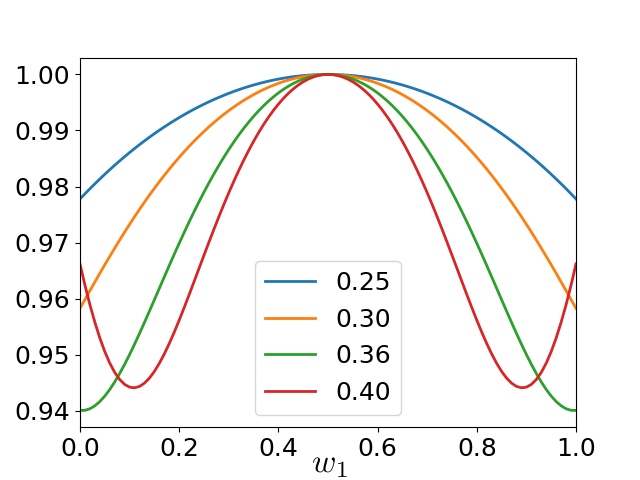}
    \includegraphics[width=0.50\textwidth]{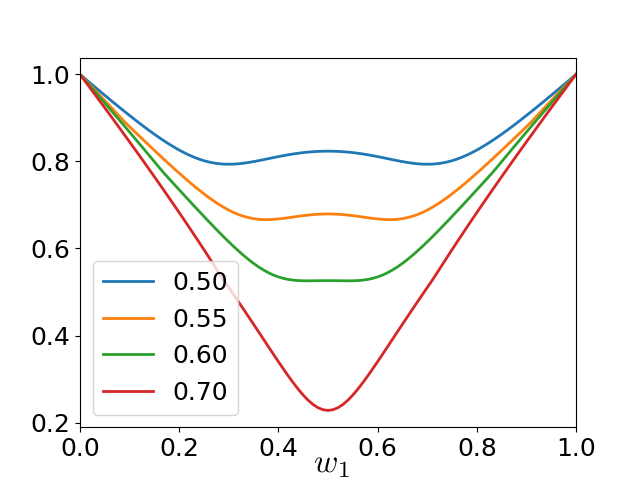}
    \hspace{-0.3\textwidth}
\caption{Left: Functional $Q(w_1,w_2)$, $w_1+w_2=1$ for different $v \in [0.25,0.4]$, normalized to the maximum value of the functional. Parameters: $\mathbf{R_0}=4, k=0.01, l=0.01, I=0.01N$.\protect\newline
    Right: Functional $Q(w_1,w_2)$, $w_1+w_2=1$ for different $v \in [0.5,0.7]$, normalized to the maximum value of the functional. Parameters: $\mathbf{R_0}=4, k=0.01, l=0.01, I=0.01N$.}
    \label{fig:13-14}
\end{figure}

Let the migration activity of infectives increase 10 times in both centers. In this case $v^*_1 \approx 0.358$ (see Figure~\ref{fig:13-14} (left)).
Note that the threshold $v^*_1$ has decreased comparing to the basic example.
Thus, in the case of high mobility of the infectives it is more reasonable to divide the vaccine between both centers.

\subsection{Non-symmetric SIR centers}\label{sec:asym}

\subsubsection{Source of the disease: different numbers of initial infectives $I_{i}(0)$}
\begin{figure}[H]
    \centering
    \hspace{-0.3\textwidth}
    \includegraphics[width=0.50\textwidth]{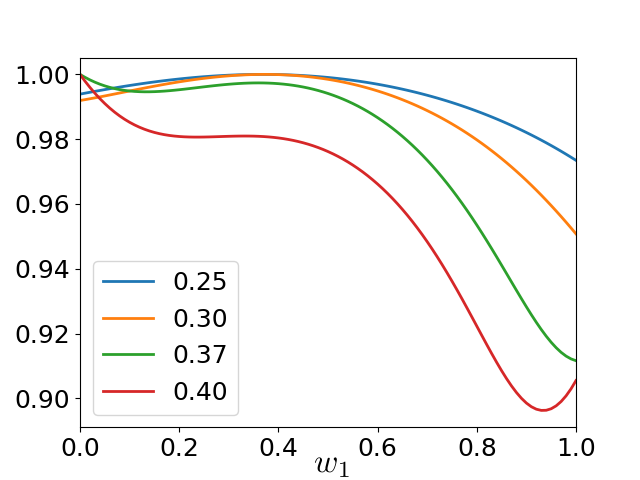}
    \includegraphics[width=0.50\textwidth]{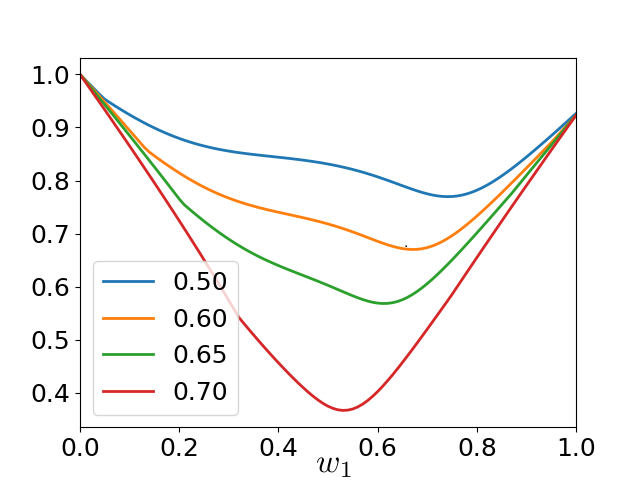}
    \hspace{-0.3\textwidth}
\caption{
	Left: Functional $Q(w_1,w_2)$, $w_1+w_2=1$ for different $v \in [0.25,0.4]$, normalized to the maximum value of the functional. Parameters: $\mathbf{R_0}=4, I_1(0)=0.01N, I_2(0)=0.1N$.\protect\newline
	Right: Functional $Q(w_1,w_2)$, $w_1+w_2=1$ for different $v \in [0.5,0.7]$, normalized to the maximum value of the functional. Parameters: $\mathbf{R_0}=4, I_1(0)=0.01N, I_2(0)=0.1N$.}
    \label{fig:15-16}
 \end{figure}

Now we study the dependence of the functional $Q$ on different numbers of initial infectives (see Figure~\ref{fig:15-16}). A common sense suggests that the entire vaccine stock is to be concentrated in the center with the greater number of infected. However, this intuition is misleading.

Surprisingly, we must allocate the most part of the vaccine in the center with the smaller number of  initial  infectives.
Center with the larger number of infectives may be interpreted as a similar center with basic parameters at some time moment $t > 0$. As shown before, the optimal vaccination time is $t^* = 0$, so it is more optimal to vaccinate the center with lower number of infectives. Eventually, having large amount of vaccine we must allocate it equally between the two centers.

\subsubsection{Mass Migration: different mobility rates $k_{ij}$, $l_{ij}$}

Let the rates $k_{12} = 10 k_{21}$ and $l_{12} = 10 l_{21}$ (i.e., the flow is directed from the first host center to the guest center).

\begin{figure}[H]
    \centering
    \hspace{-0.3\textwidth}
    \includegraphics[width=0.50\textwidth]{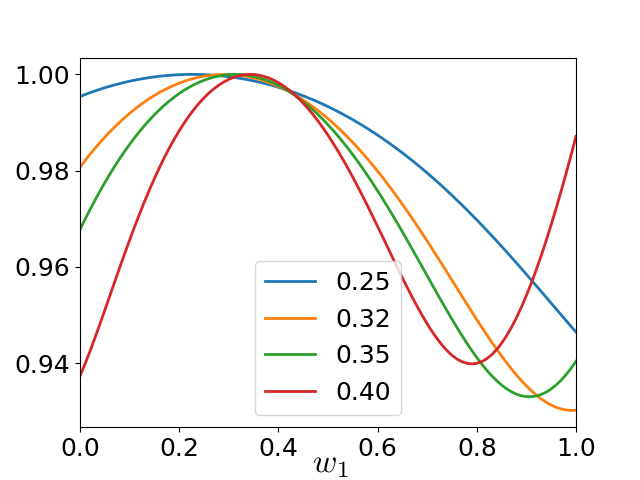}
    \includegraphics[width=0.50\textwidth]{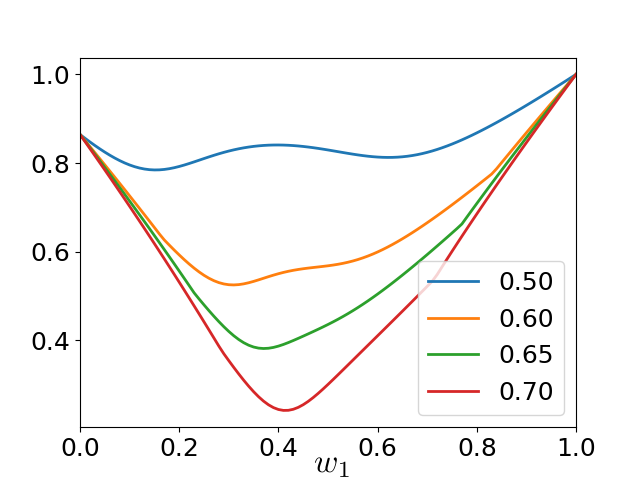}
    \hspace{-0.3\textwidth}
    \caption{
    	Left: Functional $Q(w_1,w_2)$, $w_1+w_2=1$ for different $v \in [0.07,0.4]$, normalized to the maximum value of the functional. Parameters: $\mathbf{R_0}=4, k_{12}=0.1, l_{12}=0.01, k_{21}=0.01, l_{21}=0.001, I_1(0)=I_2(0)=0.01N$.\protect\newline
    	Right: Functional $Q(w_1,w_2)$, $w_1+w_2=1$ for different $v \in [0.5,0.7]$, normalized to the maximum value of the functional. Parameters: $\mathbf{R_0}=4, k_{12}=0.1, l_{12}=0.01, k_{21}=0.01, l_{21}=0.001, I_1(0)=I_2(0)=0.01N$.}
    \label{fig:17-18}
    \end{figure}

Surprisingly, the most efficient strategy is to vaccinate the host center in case of low vaccine stock $v \leq v^*_1 \approx 0.318$ (see Figure~\ref{fig:17-18}). Moreover, a new threshold $\bar{v} \approx 0.397$ appears that correspond to a switch of vaccination from the host to the guest center. Therefore, we allocate the vaccine in the host center for $v \leq v^*_1$, divide between two centers (prioritizing host center) for $v^*_1 < v < \bar{v}$,
 and divide between two centers (prioritizing guest center) for $v > \bar{v}$.

\subsubsection{Big and Small: different number of populations}
Let the second center have the larger population: $N_2 = 2 N_1, N_1= 1000$   $I_1(0) =  I_2(0) = 0.005(N_1 + N_2)$.

\begin{figure}[H]
    \centering
    \hspace{-0.3\textwidth}
    \includegraphics[width=0.50\textwidth]{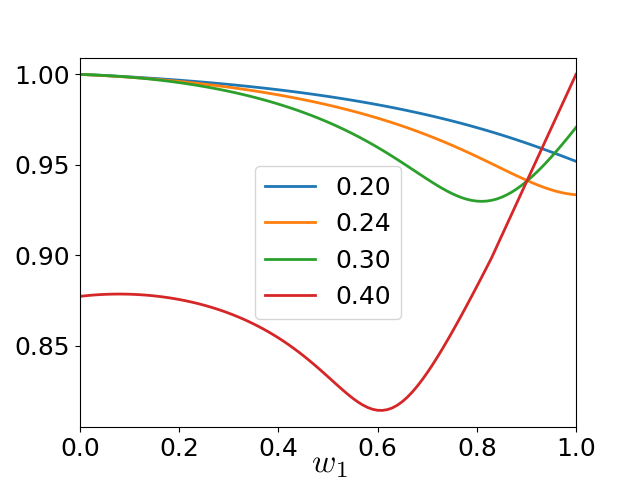}
    \includegraphics[width=0.50\textwidth]{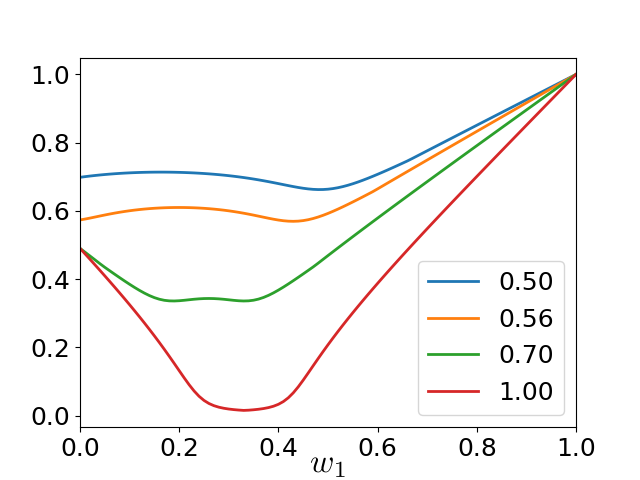}
    \hspace{-0.3\textwidth}
    \caption{
    	Left: Functional $Q(w_1,w_2)$, $w_1+w_2=1$ for different $v \in [0.20,0.40]$, normalized to the maximum value of the functional. Parameters: $\mathbf{R_0}=4, N_2=2 N_1, k_{12}=0.01, l_{12}=0.001, k_{21}=0.01, l_{21}=0.001, I_1(0)=0.005(N_1 + N_2),  I_2(0)=0.005(N_1 + N_2)$.\protect\newline
    	Right: Functional $Q(w_1,w_2)$, $w_1+w_2=1$ for different $v \in [0.5, 1.0]$, normalized to the maximum value of the functional. Parameters: $\mathbf{R_0}=4, N_2=2 N_1, k_{12}=0.01, l_{12}=0.001, k_{21}=0.01, l_{21}=0.001, I_1(0)=0.005(N_1 + N_2),  I_2(0)=0.005(N_1 + N_2)$.}
    \label{fig:19-20}
    \end{figure}

In case of small $v$ we allocate all vaccine stock in the center with the smallest population (see Figure~\ref{fig:19-20}).
The threshold is approximately $v^*_1 \approx 0.24$.

Note that the full stock $v = 1$ allows vaccinating all susceptibles. Thus, at $v = 1$ the vaccine share in smaller center is $w_1 \approx 0.33$, since we have $S_1(0) = 985$, $S_2(0) = 1985$, and we can vaccinate all population. Therefore, $w_1 = \frac{S_1}{S_1+S_2} \approx 0.33$.

\section{Conclusion}
In this paper we analyze the importance of the optimal allocation of a vaccine stockpile in order to minimize a natural functional $Q$, i.e., the  number of lost working days for the whole period of epidemic outbreak. The numerical simulations demonstrated that the optimal allocation of vaccine may reduce the functional $Q$ significantly.
The effect of optimization is not negligible: in some cases $Q$ decreases more than 2 times from the initial value.
For real life it is a fantastic result.

Further studies may be associated with the development of a model for the $n$ centers and more precise specification of interaction.
More detailed model of migration fluxes taking into account the different dynamics of susceptibles and infected species was proposed in \cite{l10}. We expect more precise resemblance to real life optimal vaccination in the model described in  \cite{l10}. Quarantine, the latent period of the disease, risk group population structure  will make a model very close to reality.

Also, the comparison of deterministic model with the stochastic SIR model will provide an additional insight of taking into account the nonzero probability that infection might fail to spread in a network of interacting centers.

Another interesting subject of research is the choice of the functional specification, i.e., it can be selected based on the price of the vaccine and economic losses related to the non-used amount of vaccine, etc.

In the future papers all these questions will be investigated and results will be compared with the real data.

\section{Acknowledgements}
The article was prepared within the framework of the Academic Fund Program at
the National Research University Higher School of Economics (HSE) and supported within the subsidy granted to the HSE by the Government of the Russian Federation for the implementation of the Global Competitiveness Programme.

\end{document}